\documentclass[aps,
groupedaddress,
nofootinbib,twocolumn]{revtex4}
\usepackage{amssymb,amsmath}
\usepackage{graphicx}
\usepackage[english]{babel}
\usepackage[usenames,dvipsnames]{color}

\newcommand{\ket}[1]{\left | #1 \right \rangle}
\newcommand{\bra}[1]{\left \langle #1 \right |}
\newcommand{\braket}[2]{\langle#1|#2\rangle}
\newcommand{\proj}[1]{\ket{#1}\!\bra{#1}}
\newcommand{\avg}[1]{\langle#1\rangle}
\newcommand{\bigavg}[1]{\big\langle#1\big\rangle}

\definecolor{darkred}{rgb}{.8,0,0}

\definecolor{darkblue}{rgb}{0,0,.7}

\newcommand{\mat}[1]{\mathsf{#1}} 

\begin{document}

\title{Local time of random walks on graphs}

\author{V\'{a}clav Zatloukal}
\email{zatlovac@gmail.com}
\homepage{zatlovac.eu}

\affiliation{Department of Physics, Faculty of Nuclear Sciences and Physical Engineering,\\
Czech Technical University in Prague, B\v{r}ehov\'{a} 7, 115 19 Praha 1, Czech Republic}

\begin{abstract}
We investigate the local (or occupation) time of a discrete-time random walk on a generic graph, and present a general method for calculating sample-path averages of local time functionals in terms of the resolvent of the transition matrix.
\end{abstract}

\maketitle

\section{Introduction}

The notion of local (or occupation) time of a stochastic process was introduced by L\'{e}vy \cite{Levy1939} as a measure of time that a stochastic trajectory spends in the vicinity of a given point in state space.
Among continuous stochastic processes, most attention has been given to the Brownian motion on a real line, whose local time was investigated, e.g., in Refs. \cite{MarcusRosen,Borodin1989,Luttinger1983}.
In Ref.~\cite{Zatloukal2017} we studied the so called L\'{e}vy random walks, which generalize the Brownian motion, and whose characteristic feature is the polynomial decay of transition functions (the heavy tails). 

In the present article we will be concerned with discrete state spaces, in which case the random walks are commonly referred to as \emph{Markov chains} \cite{Norris}. 
These are used as statistical models of various, effectively memoryless, real-world processes \cite{MeynTweedie}, such as the growth of populations, or can be employed, for example, to design efficient sampling algorithms \cite{Lovasz1993}. 
For an overview of physics-related results on random walks on graphs see, e.g., Ref.~\cite{Burioni2005}.

The local time of a discrete random walk with discrete time evolution is simply the number of times the walker visits a given location \cite{Revesz2005,Vallander1984,Sericola2000}.
For continuous time evolution, the local time is the total amount of time the walker spends at a given location \cite{Sericola2000,Brydges2007,Vallander1984II}.
(In this article only discrete time evolution will be considered.)


To be specific, let $V$ denote a set of vertices, the state space, and $E$ a set of edges that describes possible movements of the walker. Together they form a graph (or network) $G=(V,E)$. In general, we consider oriented (or directed) edges.\footnote{An unoriented edge can be realized as a composition of two edges in opposite directions.}
To each edge $(v,v')$ we assign a strictly positive number, which represents the probability that the walker localized at a vertex $v$ moves to vertex $v'$. These transition probabilities, denoted $p_{vv'}$, must satisfy the normalization conditions
\begin{equation} \label{StochMat}
\sum_{v'\in V} p_{vv'}
= 1 
\quad,\quad
\forall v \in V .
\end{equation}
The ensuing matrix $\mat{P} = (p_{vv'})$ is \emph{stochastic}, and is referred to as the \emph{transition matrix}.

The probability that the walker is found at position $v_b$ after $n$ steps of the walk, starting with certainty at position $v_a$, is given by the corresponding matrix element of the $n$-th power of the transition matrix $\mat{P}$:
\begin{equation} \label{TransProb}
\bra{v_a} \mat{P}^{n} \ket{v_b}
\equiv (\mat{P}^n)_{v_a v_b} .
\end{equation}
Here, and in the following, we use the Dirac's braket notation \cite{Dirac1939}, defining $\ket{v}$ the column vector with all entries $0$, except for the $v$-the entry, which is equal to $1$; and denoting by $\bra{v} = \ket{v}^T$ the dual row vector. We believe that this notation, common in quantum theory, proves practical also in the field of stochastic processes. (A brief introduction into the bra-ket formalism is provided in Appendix~\ref{sec:AppBraket}.)

The $n$-step transition probability of Eq.~\eqref{TransProb} can be visualized as a sum over all paths of length $n$ that start at $v_a$, and end at $v_b$. This sum is a discrete analogue of the Wiener integral of continuous stochastic processes \cite{Wiener1923} (which was a precursor of the Feynman's path integral approach to quantum mechanics \cite{FeynmanHibbs}). Indeed, the present work is inspired by the path-integral techniques, which were already successfully applied to investigate the local times of continuous random walks in Ref.~\cite{Zatloukal2017}.

The article is organized as follows. In Sec.~\ref{sec:LT} we provide the precise definition of the local time for a given sample trajectory, and study its statistical properties. For this purpose we cast sample-path averages of generic functionals of the local time (using the method of source potentials and generating functionals) in terms of the resolvent operator -- Eq.~\eqref{FuncAvgz}. In Sec.~\ref{sec:Corr} we derive formulas for the mean local time spent at a given site, and for correlations of local times between two sites, and study their large-time asymptotics (Sec.~\ref{sec:Asymp}). In Sec.~\ref{sec:OnePoint} we obtain more detailed information about the local time at a given vertex (namely, its distribution function) by investigating one-point functionals. 

Sec.~\ref{sec:Examples} is dedicated to examples of graphs, for which the resolvent, a key object in our treatment, can be calculated in closed form. We treat the complete graph, the star graph, and the (infinite) discrete line.


%
%
%

\section{Local time of a discrete random walk}
\label{sec:LT}

The (discrete-)path-integral representation of the transition probabilities of Eq.~\eqref{TransProb} is obtained easily --- by expanding the matrix multiplications. This introduces $n-1$ summation indices $v_1,\ldots,v_{n-1}$, which form, together with the initial vertex $v_a = v_0$, and the final vertex $v_b = v_n$, sample paths $\vec{v}=(v_0,v_1,\ldots,v_n)$. We have
\begin{equation}
\bra{v_a} \mat{P}^{n} \ket{v_b}
= \sum_{\substack{\vec{v} \\ v_0=v_a}}^{v_n=v_b} w(\vec{v}) ,
\end{equation}
where the weight function (i.e., the probability of a path $\vec{v}$) reads 
\begin{equation}
w(\vec{v}) 
= p_{v_0 v_1} p_{v_1 v_2} \ldots p_{v_{n-1} v_n} .
\end{equation}

\begin{figure}
\includegraphics[scale=0.8]{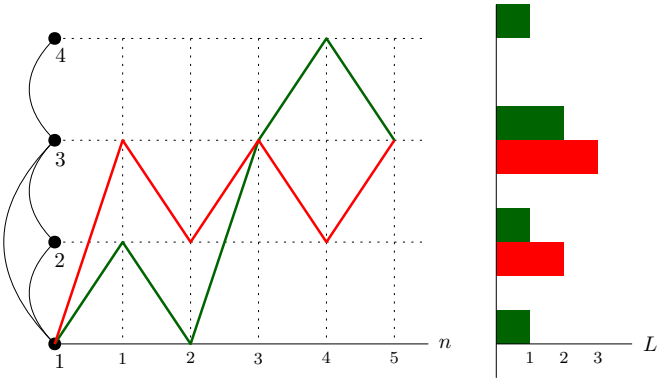}  
\caption{Sample paths of a discrete random walk on an (undirected) $4$-vertex graph, and the corresponding local time profiles after $n=5$ time-steps.}
\label{fig:paths}
\end{figure}

For a given sample trajectory $\vec{v}$, the local time at vertex $v$ is defined
\begin{equation} \label{LTdef}
L(v;n,\vec{v})
= \sum_{m=1}^n \delta_{v, v_m} .
\end{equation}
(We shall often write $L(v)$ for brevity.)
The function (or, the local time profile) $L(v)$ counts the number of times the vertices of the trajectory $\vec{v}$ coincide with $v$ (see Fig.~\ref{fig:paths} for an illustration). Note that the initial position is not included in the counting, so the sum $\sum_{v \in V} L(v)$ equals the total number of steps $n$. 

The dependence of $L(v)$ on the stochastic trajectory $\vec{v}$ implies that it is itself a random function (a family of random variables indexed by $v$). In order to quantify the local time of a random walk, we shall be therefore interested in ensemble (or sample-path) averages of functionals $F[L(v)]$ of the local time profiles. These functionals include, for example, $L(v_1)$, $L(v_1) L(v_2)$, $\ldots$, which yield the moments of the local time distribution functions, or $\delta_{L_1,L(v_1)}$, $\delta_{L_1,L(v_1)} \delta_{L_2,L(v_2)}$, $\ldots$, which yield the distribution functions themselves.

The (unnormalized) average (or expectation value) of a functional $F[L(v)]$ can be expressed as the discrete path integral
\begin{equation} \label{FuncAvgDef}
\avg{F[L(v)]}
= \sum_{\substack{\vec{v} \\ v_0=v_a}}^{v_n=v_b} w(\vec{v}) F[L(v)] .
\end{equation}
Ultimately, this quantity should be normalized by the factor $\avg{1}$ in order to possess the usual probabilistic interpretation. The normalization factor depends on the set of trajectories considered. In Eq.~\eqref{FuncAvgDef} both the initial point, and the final point are fixed, and, in general, $\avg{1} = \bra{v_a} \mat{P}^{n} \ket{v_b} \neq 1$, but we will also consider ensembles of paths with final point unspecified. In the latter case we denote ensemble averages by $\avg{\ldots}^*$, and observe that $\avg{1}^* = 1$ in consequence of the normalization conditions, Eq.~\eqref{StochMat}, for the transition matrix $\mat{P}$.

The method we now use to manipulate the right-hand side of Eq.~\eqref{FuncAvgDef} is inspired by the quantum-field-theoretical method of sources and generating functionals (see, e.g., the monograph \cite{Ramond}), and is entirely analogous to the approach used in Ref.~\cite{Zatloukal2017}.

We start by introducing an auxiliary function (the \emph{source potential}) $U:V \rightarrow \mathbb{R}$, and observe that, by definition~\eqref{LTdef}, for any path $\vec{v}$,
\begin{equation}
\sum_{m=1}^n U(v_m) 
= \sum_{v \in V} U(v) L(v) .
\end{equation}
Rather than being a specific function, the source potential should be understood as a collection of variables $U(v)$, indexed by $v$, with respect to which we can differentiate (setting $U(v)=0$, $\forall v \in V$, at the end of the calculation), so as to generate a desired functional $F[L(v)]$:
\begin{align} 
\avg{F[L(v)]}
&= F\left[ \frac{\partial}{\partial U(v)} \right]
\sum_{\substack{\vec{v} \\ v_0=v_a}}^{v_n=v_b} \!\!\!w(\vec{v})
\left. \exp \left(\sum_{m=1}^n \!U(v_m) \right) \right|_{U=0}
\nonumber\\
&= \left.F\left[ \frac{\partial}{\partial U(v)} \right]
\bra{v_a} (\mat{P} e^{\mat{U}})^n \ket{v_b} \right|_{U=0}.
\end{align}
In the second line we have introduced the diagonal matrix $\mat{U} = \sum_{v \in V} U(v) \proj{v}$, which conveniently assembles the values of the source potential, and allows to cast the result in a compact form, exhibiting the source potential as a `deformation' of the transition matrix $\mat{P}$.
Note that the quantity $\avg{F[L(v)]}$ depends on time $n$ (as well as on the initial and final positions $v_a$ and $v_b$), although this dependence is not displayed explicitly in our notation.

In the next step we pass from the discrete time variable $n$ to a continuous variable $z$ via the so-called (unilateral) $z$-transform \cite{Graf}, \cite[Ch.\,18]{Gradshteyn}. That is, we convert time series in $n$ into (their generating) functions of a new variable $z$. The functional averages in the $z$-space read
\begin{align} \label{FuncAvgz}
\avg{F[L(v)]}_z
&= \sum_{n=0}^{\infty} \avg{F[L(v)]} z^{-n}
\nonumber\\
&= -z \left.F\left[ \frac{\partial}{\partial U(v)} \right]
\bra{v_a} \mat{R}_U(z) \ket{v_b} \right|_{U=0} ,
\end{align}
where
\begin{equation}
\mat{R}_U(z) \equiv (\mat{P} e^{\mat{U}} - z\, \mat{I})^{-1}
\end{equation}  
is the \emph{resolvent} corresponding to the matrix $\mat{P} e^{\mat{U}}$, and $\mat{I}$ denotes the identity matrix. The $z$-transform can be viewed as a discrete analogue of the Laplace transform.

To pass back to the time domain one may expand the function of $z$, $\avg{F[L(v)]}_z$, in powers of $1/z$, or use contour integration techniques in the complex $z$-plane.
In addition, there exist asymptotic formulas, which relate the large-$n$ behaviour with the limit $z \rightarrow 1$ (see Appendix~\ref{sec:AppAsymp}).

To proceed further we need to specify the local time functional $F[L(v)]$. In the following, we will consider the cases $L(v_1)$ and $L(v_1)L(v_2)$, where $v_1$ and $v_2$ are some fixed vertices; and the case $f(L(v_1))$, where $f$ is specified as $f(L(v_1)) = \delta_{\ell,L(v_1)}$. (Hence $\ell$ plays a role of the variable of the ensuing local time probability distribution at a given position $v_1$.)

%
%
%

\subsection{Mean and correlations}
\label{sec:Corr}

First, let us take $F[L(v)] = L(v_1)$, and calculate
\begin{align} \label{LTFirstMom}
\avg{L(v_1)}_{z}
&= -z \left. \frac{\partial}{\partial U(v_1)}
\bra{v_a} (\mat{P} e^{\mat{U}} - z\, \mat{I})^{-1} \ket{v_b} \right|_{U=0}
\nonumber\\
&= z \bra{v_a} (\mat{P} - z\, \mat{I})^{-1} \mat{P} \proj{v_1} (\mat{P} - z\, \mat{I})^{-1} \ket{v_b}
\nonumber\\
&= z \bra{v_a} \mat{R} \mat{P} \proj{v_1} \mat{R} \ket{v_b} ,
\end{align} 
where we have made use of the differential identity $\frac{\partial \mat{A}^{-1}}{\partial u} = - \mat{A}^{-1} \frac{\partial \mat{A}}{\partial u} \mat{A}^{-1}$ for matrix-valued functions $\mat{A}(u)$. 

The result of Eq.~\eqref{LTFirstMom} is formulated in terms of the ``free" resolvent $\mat{R} \equiv \mat{R}_{U=0} = (\mat{P} - z\,\mat{I})^{-1}$, which is essentially the random walk's generating function \cite[Ch.\,2]{RudnickGaspari}. It simplifies further if we consider paths with unspecified final point. This amounts to a summation over $v_b$, that is, to a replacement of $\ket{v_b}$ by $\ket{\bf 1} \equiv (1,1,\ldots,1)^T$. Since $\mat{P} \ket{\bf 1} = \ket{\bf 1}$, due to Eq.~\eqref{StochMat}, we obtain
\begin{equation}
\bra{v_1} \mat{R} \ket{\bf 1} 
= \frac{1}{1-z} ,
\end{equation}
and hence
\begin{align} \label{LTFirstMomFreeZ}
\avg{L(v_1)}_{z}^*
&= \frac{z}{1-z} \bra{v_a} \mat{R} \mat{P} \ket{v_1} .
\end{align}

One may expand the right-hand side in powers of $1/z$ to obtain the mean local time at position $v_1$ after $n$ steps in the form of a sum of powers of the transition matrix,
\begin{equation} \label{LTFirstMomFree}
\avg{L(v_1)}^*
= \sum_{m=1}^n \bra{v_a} \mat{P}^m \ket{v_1} .
\end{equation}
Note that the same result can be obtained directly, without the resolvent method, by plugging the functional $F[L(v)] = L(v_1)$ into Eq.~\eqref{FuncAvgDef}, using the definition of the local time, Eq.~\eqref{LTdef}, and casting the ensuing discrete path integral in terms of the transition matrix $\mat{P}$. 
However, the resolvent method is more flexible, and easier to implement for more complicated functionals.

Next, consider the functional $F[L(v)] = L(v_1) L(v_2)$, which captures correlations of the local times at points $v_1$ and $v_2$. Analogous calculations as in the previous case yield 
\begin{align} \label{LTSecondMom}
\avg{L(v_1) L(v_2)}_z
= &-z^2 \bra{v_a} \mat{R} \mat{P} \ket{v_1} \bra{v_1} \mat{R} \ket{v_2} \bra{v_2} \mat{R} \ket{v_b}
\nonumber\\
&-z \bra{v_a} \mat{R} \mat{P} \ket{v_2} \bra{v_2} \mat{R} \mat{P} \ket{v_1} \bra{v_1} \mat{R} \ket{v_b} 
\end{align}
for fixed final point $v_b$, and
\begin{align} \label{LTSecondMomFree}
\avg{L(v_1) L(v_2)}_z^*
= &\frac{z^2}{z-1} \bra{v_a} \mat{R} \mat{P} \ket{v_1} \bra{v_1} \mat{R} \ket{v_2} 
\nonumber\\
&+ \frac{z}{z-1} \bra{v_a} \mat{R} \mat{P} \ket{v_2} \bra{v_2} \mat{R} \mat{P} \ket{v_1} 
\end{align}
for free final point.

\subsection{One-point functionals}
\label{sec:OnePoint}

Suppose the functional $F$ depends on the value of local time only at a single point $v$. It is then sufficient to consider the source potential that is zero everywhere except at this vertex:
\begin{equation}
\mat{U}_v = u \proj{v}
\quad,\quad u \in \mathbb{R} .
\end{equation}
In this case $e^{\mat{U}_v} = \mat{I} + (e^u-1)\proj{v}$, and we can express the ``full" resolvent in terms of the ``free" resolvent as
\begin{align} \label{ResOnePoint}
\mat{R}_{U}
&= \big(\mat{P} - z\,\mat{I} + (e^u-1) \mat{P} \proj{v}\big)^{-1} 
\nonumber\\
&= \big(\mat{I} + (e^u-1) \mat{R} \mat{P} \proj{v}\big)^{-1} \,\mat{R}
\nonumber\\
&= \mat{R} + 
\frac{(1-e^u) \mat{R P} \proj{v} \mat{R}}{1 - (1-e^u) \bra{v} \mat{R P} \ket{v}} .
\end{align}

To be more specific, let us choose the one-point functional 
\begin{equation}
\delta_{\ell,L(v)}
= \frac{1}{2\pi} \int_{0}^{2\pi} \!\!e^{i \varphi (\ell - L(v))} d\varphi ,
\end{equation}
which gives $1$ for trajectories with local time at $v$ equal to $\ell$, and $0$ otherwise. The sample-path average $\avg{\delta_{\ell,L(v)}}$ is the (unnormalized) distribution function (in variable $\ell \geq 0$) of the local time at point $v$.
Eq.~\eqref{FuncAvgz} now reads
\begin{align} \label{OnePointDistr}
\avg{\delta_{\ell,L(v)}}_z 
&= -\frac{z}{2\pi} \int_{0}^{2\pi} \!\!d\varphi \, e^{i \varphi \ell} 
\bra{v_a} \mat{R}_U(z) \ket{v_b}|_{u=-i\varphi} ,
\end{align}
so substituting from Eq.~\eqref{ResOnePoint}, using Formula~\eqref{Int1} together with \eqref{Int1alpha}, and the fact that $\mat{R} \mat{P} = \mat{I} + z\mat{R}$, we obtain
\begin{align} \label{OnePointDistr1}
\avg{\delta_{\ell,L(v)}}_z 
&= -z \bra{v_a} \mat{R} \ket{v_b} \delta_{\ell,0}
\nonumber\\
&+ \frac{\bra{v_a} \mat{R P} \proj{v} \mat{R} \ket{v_b}}{\bra{v} \mat{R P} \ket{v}}
\left( z\, \delta_{\ell,0} + \frac{\bra{v} \mat{R P} \ket{v}^{\ell}}{z^\ell \bra{v} \mat{R} \ket{v}^{\ell+1}} \right) .
\end{align}
This is the most general formula for the one-point local time distribution in the $z$-domain.

A more compact result is obtained when we focus on the distribution at the origin of the walk ($v = v_a$), and do not specify the final point ($\ket{v_b} = \ket{\bf 1}$):
\begin{equation} \label{OnePointDistr2}
\avg{\delta_{\ell,L(v_a)}}_z^*
= \frac{\bra{v_a} \mat{R P} \ket{v_a}^{\ell}}{(1-z) z^\ell \bra{v_a} \mat{R} \ket{v_a}^{\ell+1}} .
\end{equation} 

A remarkably simple result is also obtained for $\ell=0$ (and $v$ arbitrary), i.e., for the probability that the walker, during the steps $1,\ldots,n$, never visits a site $v$:
\begin{equation} \label{ZeroLT}
\avg{\delta_{0,L(v)}}
= \bra{v_a} (\mat{P}-\mat{P}\proj{v})^n \ket{v_b} ,
\end{equation}
where $\mat{P}-\mat{P}\proj{v}$ is the matrix $\mat{P}$ with $v$-th column set to zero (which effectively assigns zero weights to all paths entering the vertex $v$). To verify this result, we observe that for $\ell=0$ Eq.~\eqref{OnePointDistr1} reduces to
\begin{equation}
\avg{\delta_{0,L(v)}}_z 
= -z \bra{v_a} \Big(\mat{I} - \frac{\mat{R}\mat{P}\proj{v}}{z\bra{v}\mat{R}\ket{v}} \Big) \mat{R} \ket{v_b} ,
\end{equation}
and that the same expression results by $z$-transforming Eq.~\eqref{ZeroLT}.
$\avg{\delta_{0,L(v)}}$ is in fact the initial condition for recurrence relations that allow to determine the quantities $\avg{\delta_{\ell,L(v)}}$ also for $\ell \geq 0$ \cite[Sec.\,3.1]{Sericola2000}.

In this article we do not consider the general problem of functionals that depend on the local time at multiple points, 
although such analysis, along the lines of Appendix~A in \cite{Zatloukal2017} or Ref.~\cite{Vallander1984}, is certainly possible. Alternatively, one can use the recurrence relations of Ref.~\cite[Sec.\,4.1]{Sericola2000}.

\subsection{Large-time asymptotics}
\label{sec:Asymp}

In general, the transition from the $z$-domain averages $\avg{\ldots}_z$ to the time-domain averages $\avg{\ldots}$ is a hard task, since the free resolvent $\mat{R}$ can be a complicated function of $z$. 
In this subsection we focus on the regime when the number of time steps $n$ goes to infinity, and make use of the final value theorem of Appendix~\ref{sec:AppAsymp}, which relates this limit to the limit of $z$ approaching $1$ from above. 

We will assume that the transition matrix $\mat{P}$ corresponds to a finite strongly connected underlying graph, meaning that there is a path between any two vertices in both directions. Since $\mat{P}$ is stochastic, it has a (right) eigenvector $\ket{\bf 1}$ corresponding to eigenvalue $1$, and, by the Perron-Frobenius theorem \cite[Ch.\,8.8]{GodsilRoyle}, this eigenspace is one-dimensional, and all other eigenvalues have magnitude $\leq 1$.

The invariant (or stationary) distribution $\bra{\pi}$ is the left eigenvector of $\mat{P}$ with eigenvalue $1$, normalized so that $\braket{\pi}{{\bf 1}} = 1$. We can split the transition matrix $\mat{P}$ as
\begin{equation}
\mat{P}
= \ket{{\bf 1}}\!\bra{\pi} + (\mat{I}-\ket{{\bf 1}}\!\bra{\pi}) \mat{P} ,
\end{equation}
where the first term is the (so-called Perron) projection on the eigenspace corresponding to the eigenvalue~$1$.

Since the product of the projector $\ket{{\bf 1}}\!\bra{\pi}$ with the projector $\mat{I}-\ket{{\bf 1}}\!\bra{\pi}$ vanishes, the free resolvent decomposes analogously,
\begin{equation}
\mat{R}
= \frac{1}{1-z} \ket{{\bf 1}}\!\bra{\pi} 
+ \frac{1}{\mat{P} - z \,\mat{I}} (\mat{I}-\ket{{\bf 1}}\!\bra{\pi}) .
\end{equation}
A crucial observation is that the second term drops out in the following limit:
\begin{equation} \label{Rlimit}
\lim_{z \searrow 1} (1-z) \mat{R}(z)
= \ket{{\bf 1}}\!\bra{\pi} .
\end{equation}

Eq.~\eqref{Rlimit} can now be used to investigate the large-$n$ behaviour of the local time. To obtain finite results, we consider the local time fraction $\frac{1}{n} L(v;n)$, i.e., the relative amount of time spent at a vertex $v$.

As an illustration, let us calculate the mean and correlations, which, in the $z$-domain, are given by formulas~\eqref{LTFirstMomFreeZ} and \eqref{LTSecondMomFree}, respectively. By Eq.~\eqref{Largen}, we find
\begin{align} \label{LargenMean}
\lim_{n \rightarrow \infty} \frac{\avg{L(v_1)}^*}{n}
&= \lim_{z \searrow 1} (z-1)^2 \avg{L(v_1)}_z^*
\nonumber\\
&= \braket{\pi}{v_1} ,
\end{align}
and
\begin{align}
\lim_{n \rightarrow \infty} \frac{\avg{L(v_1) L(v_2)}^*}{n^2}
&= \lim_{z \searrow 1} \frac{(z-1)^3}{2} \avg{L(v_1) L(v_2)}_z^*
\nonumber\\
&= \braket{\pi}{v_1} \braket{\pi}{v_2} .
\end{align}

It is worth to note that these results do not depend on the initial position $v_a$. That is, after sufficiently many time-steps the walker looses memory of where it started. Also, the covariance of local times at two points diminishes:
\begin{equation}
\lim_{n \rightarrow \infty} \frac{1}{n^2} \bigavg{\big(L(v_1)-\avg{L(v_1)}^*\big) \big(L(v_2)-\avg{L(v_2)}^*\big)}^*
= 0 .
\end{equation}
For $v_1=v_2$ this implies that for any trajectory $\vec{v}$, the local time fraction $\frac{1}{n} L(v_1;n,\vec{v})$ converges (in quadratic mean, and hence in probability \cite[Ch.\,17]{JacodProtter}) 
to the sample-path average $\frac{1}{n} \avg{L(v_1)}^*$, which in turn is given by the invariant distribution as $\braket{\pi}{v_1}$. Thus, we recover the ergodic theorem of Ref.~\cite[Ch.\,1.10]{Norris}.

In closing this section, let us remark that there is a natural (unbiased) way of assigning a transition matrix to a graph with adjacency matrix $\mat{A}$, namely,
\begin{equation} \label{TransAdj}
\bra{v} \mat{P} \ket{v'}
= \frac{\bra{v} \mat{A} \ket{v'}}{\bra{v} \mat{A} \ket{{\bf 1}}} .
\end{equation}
If the graph is undirected, and hence the adjacency matrix symmetric, the invariant distribution is given by a simple explicit formula
\begin{equation} \label{InvarDist}
\braket{\pi}{v} 
= \frac{\bra{v} \mat{A} \ket{{\bf 1}}}{\bra{{\bf 1}} \mat{A} \ket{{\bf 1}}} .
\end{equation}
Here, $\bra{v} \mat{A} \ket{{\bf 1}}$ is the number of (unoriented) edges connected with $v$ (the so-called \emph{degree} of $v$), and $\bra{{\bf 1}} \mat{A} \ket{{\bf 1}}$ is twice the total number of edges in the graph.

The degree quantifies how well a given vertex $v$ is connected within the network (it is a measure of \emph{centrality} of the vertex \cite[Ch.\,7]{Newman}). In view of Eqs.~\eqref{LargenMean} and \eqref{InvarDist}, it coincides, up to the normalization factor $\bra{{\bf 1}} \mat{A} \ket{{\bf 1}}$, with the average local time fraction at point $v$ in the limit $n \rightarrow \infty$, which bears no trace of the initial position $v_a$.
For finite times $n$, therefore, $\frac{1}{n} \avg{L(v)}^*$ can be thought of as a (time-dependent) centrality of the vertex $v$ relative to the initial position of the walk $v_a$.

\section{Examples}
\label{sec:Examples}

We provide several simple examples of (undirected) graphs to illustrate the resolvent method for calculating local times. In each example we calculate the free resolvent $\mat{R}(z)$, which can then be used to determine sample-path averages of local time functionals of interest.

\begin{figure}[h]
\includegraphics[scale=0.5]{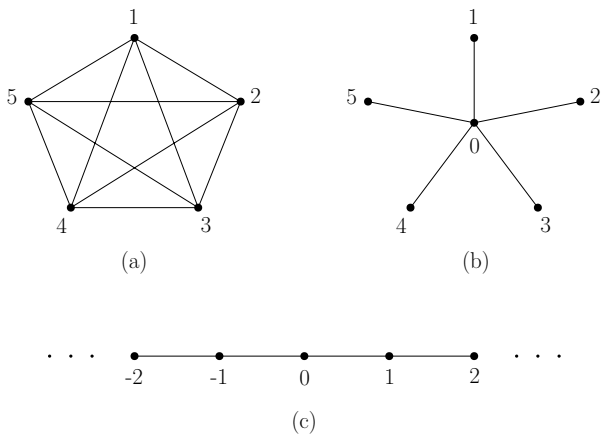}  
\caption{Examples of (undirected) graphs: (a) complete graph, (b) star graph, (c) discrete line.}
\label{fig:examples}
\end{figure}

\subsection{Complete graph}

The complete graph on $N$ vertices (Fig.~\ref{fig:examples}(a)) has the adjacency matrix
\begin{equation}
\mat{A}
= \mat{J} - \mat{I} 
\quad,\quad
\bra{v} \mat{J} \ket{v'}
= 1 \quad(\forall v,v') .
\end{equation}
Let us define the transfer matrix according to Eq.~\eqref{TransAdj},
\begin{equation}
\mat{P} 
= \frac{\mat{J} - \mat{I}}{N-1} .
\end{equation}
Since $\mat{J}^2 = N \mat{J}$, the resolvent $\mat{R} = (\mat{P} - z\,\mat{I})^{-1}$ must be a sum of two terms --- one proportional to the identity matrix $\mat{I}$, and the other proportional to the matrix $\mat{J}$. Fixing the coefficients yields
\begin{equation}
\mat{R}
= -\frac{1}{1 + z(N-1)} \left( (N-1) \,\mat{I} + \frac{\mat{J}}{z-1} \right) .
\end{equation}

Having an explicit form of the resolvent, we can calculate, for example, the mean local time for paths with unspecified final point.
According to Eq.~\eqref{LTFirstMomFreeZ} we have
\begin{equation}
\avg{L(v_1)}_{z}^*
= \frac{z}{1-z} \frac{1}{1 + z(N-1)} 
\left( \braket{v_a}{v_1} + \frac{z}{1-z} \right) ,
\end{equation}
which, upon expansion in powers of $1/z$, yields for any vertex $v_1 \neq v_a$
\begin{equation} \label{LTMeanComplete}
\avg{L(v_1)}^*
= \frac{1}{N} \left( n + \frac{1}{N} - \frac{(-1)^{n}}{N(N-1)} \right) .
\end{equation}
For the initial vertex, the mean local time is delayed by one time-step: $\avg{L(v_a)}^*|_n = \avg{L(v_1)}^*|_{n-1}$.
Note that for sufficiently large time $n$ the mean local time on any vertex grows as $n/N$.

\subsection{Star graph}

The star graph (Fig.~\ref{fig:examples}(b)) with $N$ peripheral vertices (labelled $1,\ldots,N$) situated around the central vertex (labelled $0$) defines the transfer matrix
\begin{equation}
\mat{P}
= \begin{pmatrix}
0 & \frac{1}{N} & \ldots & \frac{1}{N} \\
1 & 0 & \ldots & 0 \\
\vdots & \vdots & ~ & \vdots \\
1 & 0 & \ldots & 0
\end{pmatrix} .
\end{equation}
We may observe that $\mat{P}^3 = \mat{P}$, and so the resolvent contains at most the second power of $\mat{P}$. The coefficients of $\mat{I}$, $\mat{P}$, and $\mat{P}^2$ are readily determined, and we find
\begin{equation}
\mat{R}
= -\frac{\mat{I}}{z} + \frac{\mat{P}}{1-z^2} + \frac{\mat{P}^2}{z(1-z^2)} .
\end{equation}

Let us calculate the mean local time fraction according to Eq.~\eqref{LargenMean}:
\begin{equation}
\lim_{n \rightarrow \infty} \frac{\avg{L(v_1;n)}^*}{n}
= \frac{1}{2} \bra{v_a} \mat{P} + \mat{P}^2 \ket{v_1}
= \begin{cases}
\frac{1}{2} &{\rm if}~v_1=0 \\
\frac{1}{2 N} &{\rm if}~v_1 \neq 0
\end{cases} .
\end{equation}
The result is indeed the invariant distribution as given by Eq.~\eqref{InvarDist}. It is worth to point out that in this example the invariant distribution is not a limiting distribution of the walk, $\lim_{n \rightarrow \infty} \bra{v_a}\mat{P}^n$, as the latter does not exist. (The walker oscillates between the central vertex, and the peripheral vertices.)

\subsection{Discrete line}
\label{sec:ExDiscrLine}

Vertices of the discrete line graph are identified with integer numbers ($v \in \mathbb{Z}$), and edges are placed between nearest neighbours (see Fig.~\ref{fig:examples}(c)). The transition matrix reads
\begin{equation} \label{Pline}
\mat{P} 
= \frac{1}{2} \sum_{v \in \mathbb{Z}}
( \ket{v}\!\bra{v+1} + \ket{v+1}\!\bra{v} ) .
\end{equation}

To find the (free) resolvent $\mat{R}(z)$, we observe a generic relation $\mat{R} \mat{P} - z \mat{R} = \mat{I}$, which for $\mat{P}$ defined in Eq.~\eqref{Pline} provides a linear recurrence equation for $\mat{R}$ (in variable $v$),
\begin{equation}
\frac{1}{2} \bra{v_a} \mat{R} \ket{v+1} 
+ \frac{1}{2} \bra{v_a} \mat{R} \ket{v-1} 
- z \bra{v_a} \mat{R} \ket{v}
= \braket{v_a}{v} .
\end{equation}
This has a solution (that remains finite as $v \rightarrow \pm\infty$ for all $z > 1$)
\begin{equation}
\bra{v_a} \mat{R} \ket{v}
= -\frac{(z - \sqrt{z^2-1})^{|v_a-v|}}{\sqrt{z^2-1}} .
\end{equation}

Let us calculate the local time distribution at the origin of the walk for trajectories with unspecified final point. In the $z$-domain we have, according to Eq.~\eqref{OnePointDistr2},
\begin{equation}
\avg{\delta_{\ell,L(v_a)}}_z^*
= \frac{\sqrt{z+1}}{\sqrt{z-1}}
\left(1 - \frac{\sqrt{z^2-1}}{z} \right)^\ell .
\end{equation}
Consider, for simplicity, $\ell=0$. 
The probability that the local time observed at $v_a$ is zero after $n=2m$ or $2m+1$ time-steps is
\begin{equation}
\avg{\delta_{0,L(v_a)}}^*
= \frac{1}{2^{2m}} \binom{2m}{m} ,
\end{equation}
in agreement with Thm.~9.3 in \cite{Revesz2005}. This vanishes as $n \rightarrow \infty$, i.e., with probability $1$, the walker hits the initial position during an infinitely long random walk on a discrete line, recovering the classical result of P\'{o}lya \cite{Polya1921}. 
(For return probabilities of random walks on infinite graphs and their applications see Ref.~\cite{Woess}.)

\section{Conclusion and outlook}

The local time is a natural, intuitively motivated, characteristic of a random walk. As such, we believe, it can provide a common ground for various random-walk-related problems. For instance, in Example~\ref{sec:ExDiscrLine} it was noted that the P\'{o}lya recurrence problem is a question about the amount of trajectories with local time at the initial position strictly greater than zero. 

The ballot problem can serve as another example \cite[Ch.\,III.1]{Feller}. It can be formulated as a problem to count the number of random walk trajectories on a discrete line, which start at $v_a = 0$, end at $v_b \geq 0$, and never visit negative-numbered sites; in other words, the number of trajectories whose local time at site $-1$ is zero.

In addition, the problem of self-avoidance of the random walk trajectories can be phrased as a requirement that the local time on all sites be at most $1$.

In this article we showed how to write the sample-path average of any local time functional in terms of the resolvent matrix corresponding to the transition matrix of the walk. To this end we employed a quantum-field-theory-inspired method of source potentials, and the $z$-transform, which turns sequences indexed by the time variable $n$ into (more convenient) functions of a continuous variable $z$. 

The presented method does not presume any particular form of the underlying graph. However, to obtain explicit results, one needs to  consider sufficiently simple and symmetric graphs (some simple examples were provided in Sec.~\ref{sec:Examples}), so that the resolvent can be calculated, and also sufficiently simple local time functionals, so that the $z$-transform inversion is manageable. For complicated graphs one can use approximation techniques, such as the large-time asymptotics of Sec.~\ref{sec:Asymp}.

Apart from the discrete time evolution considered here, random walks on graphs can be defined also with a continuous time variable $t$ \cite[Ch.\,2]{Norris}. The time evolution is then dictated by the matrix $e^{t \mat{Q}}$, where $\mat{Q}$ accommodates transition rates between the vertices of the graph. The method of Sec.~\ref{sec:LT} should be applicable, with appropriate modifications, also in the continuous-time domain.


\section*{Acknowledgment}

This work was supported by the Czech Science Foundation, grant number GA \v{C}R 19-15744Y.

\appendix

\section{The bra-ket notation}
\label{sec:AppBraket}

Assume, for concreteness, that the vertices of the graph $G$ are labelled by integers, so that $v = 1, \ldots, |V|$, where $|V|$ is the total number of vertices.

The `kets' $\ket{v}$, $v \in V$, denote the (column) vectors of the standard basis of the $|V|$-dimensional real vector space of functions from $V$ to $\mathbb{R}$:
\begin{equation}
\ket{v}
\equiv (\delta_{v,1},\ldots,\delta_{v,|V|})^T
= (0,\ldots,0,\underset{v}{1},0,\ldots,0)^T .
\end{equation}
The `bras' are the dual (row) vectors
\begin{equation}
\bra{v} = \ket{v}^T ,
\quad\textrm{hence}\quad
\braket{v}{v'}
\equiv \bra{v}\ket{v'} 
= \delta_{v,v'} .
\end{equation}

Matrix elements $A_{v v'}$ of a $|V| \times |V|$ matrix $\mat{A}$, and conversely the matrix $\mat{A}$ in terms of its matrix elements, are now represented, respectively, as
\begin{equation}
A_{v v'}
= \bra{v} \mat{A} \ket{v'} 
\quad,\quad
\mat{A}
= \sum_{v,v'} \ket{v} A_{v v'} \bra{v'} .
\end{equation}
For example, note that the identity matrix reads
\begin{equation}
\mat{I}
= \sum_v \proj{v} ,
\end{equation}
where $\proj{v}$ are projectors on the vertices $v = 1,\ldots,|V|$.

Let us remark that it is common to regard the scalar $\bra{v} \mat{A} \ket{v'}$ as a single bracket, and write, for brevity,
$\bra{v} \mat{A} \ket{v'}^{\ell}$
instead of $(\bra{v} \mat{A} \ket{v'})^{\ell}$ (see, e.g., Eq.~\eqref{OnePointDistr1}).

A generic vector $f$ (that is, a function $f:V\rightarrow\mathbb{R}$) is denoted by the ket $\ket{f}$, and its value at a vertex $v$ expressed as $f(v) = \braket{v}{f}$. Note then that the standard scalar product between any $f$ and $g$ can be cast as
\begin{equation}
\sum_v g(v) f(v)
= \sum_v \braket{g}{v} \braket{v}{f}
=\bra{g} \mat{I} \ket{f}
= \braket{g}{f} .
\end{equation}

\section{List of formulas}

Adapting the row 3 in Table 18.4 of Ref.~\cite{Gradshteyn}, we have, for an integer $\ell$,
\begin{equation} \label{Int1}
\int_0^{2\pi} \frac{d\varphi}{2\pi} \, 
\frac{e^{i \varphi \ell}}{1- \alpha\,(1-e^{-i\varphi})}
= \begin{cases}
-\frac{\alpha^\ell}{(\alpha-1)^{\ell+1}}  & \textrm{for}~ \ell \geq 0
\\
0 & \textrm{for}~ \ell < 0
\end{cases} ,
\end{equation}
provided that $\big| \frac{\alpha}{\alpha-1}\big| < 1$, i.e., $\alpha \in (-\infty,\frac{1}{2})$.
In our case of interest, Eq.~\eqref{OnePointDistr}, 
\begin{equation} \label{Int1alpha}
\alpha
= \bra{v} \mat{R} \mat{P} \ket{v}
= \bra{v} \frac{\frac{\mat{P}}{z}}{\frac{\mat{P}}{z} - \mat{I}} \ket{v} ,
\end{equation}
and the condition on $\alpha$ is met when $z > 1$, since all eigenvalues of the stochastic matrix $\mat{P}$ have magnitude $\leq 1$ (see Sec.~\ref{sec:Asymp}).

\section{Final value theorem}
\label{sec:AppAsymp}

Consider a sequence $(f_n)_{n=0}^\infty$, and its $z$-transform
\begin{equation} \label{Fzfn}
F(z) = \sum_{n=0}^\infty f_n z^{-n}
\quad,\quad
z > 1 .
\end{equation}
The \emph{final value theorem} (see e.g. \cite[p.\,101]{Graf}) states that 
\begin{equation}
\lim_{n \rightarrow \infty} f_n
= \lim_{z \searrow 1} (z-1) F(z) ,
\end{equation}
provided the limit on the left-hand side exists.

Suppose $f_0 = 0$ (which is the case for local time), divide Eq.~\eqref{Fzfn} by $z$, and integrate to find
\begin{equation}
(I_F)(z)
\equiv \int_z^{+\infty} \!\!dz' \frac{F(z')}{z'}
= \sum_{n=1}^\infty \frac{f_n}{n} z^{-n} .
\end{equation}
That is, the sequence $\frac{f_n}{n}$ has $z$-transform $I_F(z)$, and we can cast the finite value theorem in the form
\begin{equation}
\lim_{n \rightarrow \infty} \frac{f_n}{n}
= \lim_{z \searrow 1} \frac{I_F(z)}{(z-1)^{-1}}
= \lim_{z \searrow 1} (z-1)^2 F(z) ,
\end{equation}
where in the last step we have used the L'H\^{o}pital rule, and realized that $\frac{F(z)}{z} \sim F(z)$ in the limit $z \rightarrow 1$.
Iterating, we find a formula for generic $k \in \mathbb{N}$:
\begin{equation} \label{Largen}
\lim_{n \rightarrow \infty} \frac{f_n}{n^k}
= \lim_{z \searrow 1} \frac{(z - 1)^{k+1}}{k!} F(z) .
\end{equation}

%
%
%
%
%


\end{document}